\documentclass[aps,prd,twocolumn,superscriptaddress,showpacs,preprintnumbers]{revtex4-1}
\usepackage{amsmath,amssymb,graphics,hyperref,color,subfigure}
\hypersetup{ 
    pdfnewwindow=true,      
    colorlinks=true,       
    linkcolor=blue,          
    citecolor=blue,        
    filecolor=blue,      
    urlcolor=blue        
} 

\begin{document}

\title{Understanding the nature of $\Lambda (1405)$ through Regge physics}

\author{C\'esar Fern\'andez-Ram\'{\i}rez}\email{cesar.fernandez@nucleares.unam.mx}
\affiliation{Instituto de Ciencias Nucleares, Universidad Nacional Aut\'onoma de M\'exico,  
A.P. 70-543, Ciudad de M\'exico 04510, M\'exico}
\affiliation{Theory Center, Thomas Jefferson National Accelerator Facility,
12000 Jefferson Avenue, Newport News, VA 23606, USA}

\author{Igor V.~Danilkin}
\affiliation{Institut f\"ur Kernphysik and PRISMA Cluster of Excellence, 
Johannes Gutenberg Universit\"at, D-55099 Mainz, Germany}

\author{Vincent Mathieu}
\affiliation{Center for Exploration of Energy and Matter, Indiana University, Bloomington, IN 47403, USA}
\affiliation{Physics Department, Indiana University, Bloomington, IN 47405, USA}

\author{Adam P.~Szczepaniak}
\affiliation{Theory Center, Thomas Jefferson National Accelerator Facility,
12000 Jefferson Avenue, Newport News, VA 23606, USA}
\affiliation{Center for Exploration of Energy and Matter, Indiana University, Bloomington, IN 47403, USA}
\affiliation{Physics Department, Indiana University, Bloomington, IN 47405, USA}

\preprint{JLAB-THY-15-2182}

\collaboration{Joint Physics Analysis Center}

\begin{abstract}
It appears that there are two resonances with $J^P=  1/2^-$ quantum numbers 
in the energy region near the $\Lambda(1405)$ hyperon. 
The nature of these states is a topic of current debate.  
To provide further insight we use Regge phenomenology to access 
how these two resonances fit the established hyperon spectrum.  
We find that only one of these resonances is compatible with a three-quark state. 
\end{abstract}
\pacs{14.20.Jn}
\date{\today}
\maketitle

Baryon spectroscopy remains as one of the main tools for the investigation of strong interactions  
in Quantum Chromodynamics  (QCD). 
In the strange baryon sector, which contains $\Lambda$ and $\Sigma$ hyperons, 
the first excitation of the isospin-0 $uds$ system is  the $\Lambda(1405)$ \cite{Alston61}. 
It is approximately 300 MeV above the ground state, $\Lambda(1116)$ \cite{PDG2014}. Its 
spin and parity have recently been confirmed to be 
$J^P=  1/2^-$ \cite{Moriya2014} but its composition is still debatable 
\cite{lattice,Engel13,Hall2015,molecule,twopoles,Mai2015,debatable,quarkmodels,Santopinto15,Faustov15,LargeNc}.
Lattice QCD computations related to the $\Lambda (1405)$ 
have appeared only recently \cite{lattice,Engel13,Hall2015} and the results are inconclusive. 
For example, in  Ref.~\cite{Engel13} $\Lambda (1405)$  emerges as a three-quark 
state while in Ref.~\cite{Hall2015} it seems to be more like a $\bar{K}N$ molecule.
Although the resonant nature of the $\Lambda(1405)$ has been ignored in these calculations.
On the phenomenological side, a combined amplitude analysis of $\bar{K}N$ scattering and
$\pi \Sigma K^+$ photoproduction  
\cite{twopoles,Mai2015,molecule} finds that in the region of the $\Lambda (1405)$ there are actually 
two resonances, one located at $1429^{+8}_{-7}-i \: 12 ^{+2}_{-3}$ MeV  and the other at
$1325^{+15}_{-15}-i\: 90 ^{+12}_{-18}$ MeV \cite{Mai2015,Roca2015}. 

In this article we employ Regge analysis  \cite{PDBCollins}
to shed more light on the nature of the $\Lambda (1405)$. 
From first principles it follows that poles in partial waves are analytically connected 
by Regge trajectories \cite{Gribov}  and analytical properties of trajectories,  
\textit{e.g.}~deviations from linearity, carry imprints 
of the underlying quark-gluon dynamics \cite{Chiu,quarklinear,mesonregge}. 

To perform the  analysis of the Regge trajectories we need to know the pole positions of the
low-lying hyperons that belong to  the $\Lambda$ Regge trajectories. 
We also use the $\Sigma$ Regge trajectories as a  benchmark. 
In Table \ref{tab:poles} we list the hyperon resonances with spin up to $J =7/2$ used in this analysis.
As discussed above, the lowest two $\Lambda$  states are the ground state 
$\Lambda (1116)$ and the $\Lambda (1405)$. The corresponding states in the isovector  
sector are identified with the  $\Sigma (1192)$ and the $\Sigma (1385)$. 
These states anchor the four leading Regge trajectories.
The $\Lambda (1116)$,  $\Sigma (1192)$, and $\Sigma (1385)$  are well established and their parameters  
are taken  from the \textit{Review of Particle Physics} \cite{PDG2014}.
The two poles in the $\Lambda (1405)$ region that we want to study are labeled as 
$\Lambda (1405)_a$ and $\Lambda (1405)_b$. 
Their parameters are taken from Ref.~\cite{Mai2015} (see Table \ref{tab:poles}).
All remaining hyperons on the leading Regge trajectories have masses  above the $\bar{K}N$ threshold
and spin $J \geq 3/2$. Parameters of these resonances  are taken from the recent analysis of 
$\bar{K}N$ partial wave amplitudes in Ref.~\cite{FR2015},  
which is based on an analytical  coupled-channel $K$-matrix approach.   

\begin{table}
\caption{Summary of pole masses ($M_p=\Re\: \sqrt{s_p}$) 
and widths ($\Gamma_p=-2 \: \Im\: \sqrt{s_p}$) in MeV.
$I$ stands for isospin, $\eta$ for naturality, $J$ for total angular momentum, 
and $P$ for parity.
Naturality and parity are related by $\eta=\tau P$ where $\tau$ is the signature.
For baryons, $\eta=+1$, natural parity, if $P=(-1)^{J-1/2}$, and
$\eta=-1$, unnatural parity, if $P=-(-1)^{J-1/2}$. 
Errors for $\Lambda(1405)$ states have been symmetrized for the calculations. 
Errors in Ref.~\cite{FR2015} are statistical.} \label{tab:poles}
\begin{ruledtabular}
\begin{tabular}{cccccc}
$I^\eta \: J^P $ & $M_p$ &$\Gamma_p $ & Name & Status & Ref.\\
\hline
$0^-\: \frac{1}{2}^-$ & $1429(8) $&$24(6)$  & $\Lambda (1405)_a$&&\cite{Mai2015}\\
$0^-\: \frac{1}{2}^-$ & $1325(15)$&$180(36)$  & $\Lambda (1405)_b$&&\cite{Mai2015}\\
$0^-\:\frac{3}{2}^+$ & $1690.3(3.8) $&$ 46(11)$ &---&&\cite{FR2015} \\
$0^-\:\frac{5}{2}^-$ & $1821.4(4.3) $&$102.3(8.6)$ &$\Lambda (1830)$&****&\cite{FR2015}\\
$0^-\:\frac{7}{2}^+$ & $2012(81) $&$210(120) $ &$\Lambda (2020)$&*&\cite{FR2015}\\
\hline
$0^+\:\frac{1}{2}^+$ & $1116$&$0$  & $\Lambda (1116)$&****&\cite{PDG2014} \\
$0^+\:\frac{3}{2}^-$ & $1519.33(34)$&$17.8(1.1)$ & $\Lambda (1520)$&****&\cite{FR2015}\\
$0^+\:\frac{5}{2}^+$ & $1817(57)$&$  85(54) $ & $\Lambda (1820)$&****&\cite{FR2015}\\
$0^+\:\frac{7}{2}^-$& $2079.9(8.3) $&$216.7(6.8) $& $\Lambda (2100)$&****&\cite{FR2015}\\
\hline
$1^- \: \frac{3}{2}^+$ & $1385(2) $&$37(5)$&$\Sigma (1385)$&****&\cite{PDG2014} \\
$1^- \: \frac{5}{2}^-$ & $1744(11)$&$165.7(9.0) $  & $\Sigma (1775)$&****&\cite{FR2015}\\
$1^- \: \frac{7}{2}^+$ & $2024(11)$&$189.5(8.1)$ & $\Sigma (2030)$&****&\cite{FR2015}\\
\hline
$1^+ \: \frac{1}{2}^+$  & $1192$&$0$& $\Sigma (1192)$&****&\cite{PDG2014}\\
$1^+ \: \frac{3}{2}^-$ & $1666.3(7.0)$&$26(19)$  &$\Sigma (1670)$&****&\cite{FR2015}\\
$1^+ \: \frac{5}{2}^+$ & $1893.9(7.2)$&$59(42)$& $\Sigma (1915)$&**** &\cite{FR2015}\\
$1^+ \: \frac{7}{2}^-$ & $2177(12)$&$156(19)$  & $\Sigma (2100)$&*&\cite{FR2015}\\
\end{tabular}
\end{ruledtabular}
\end{table}

\begin{figure}
\centering
\subfigure[\ $\Lambda$ resonances.]{
\rotatebox{0}{\scalebox{0.3}[0.3]{\includegraphics{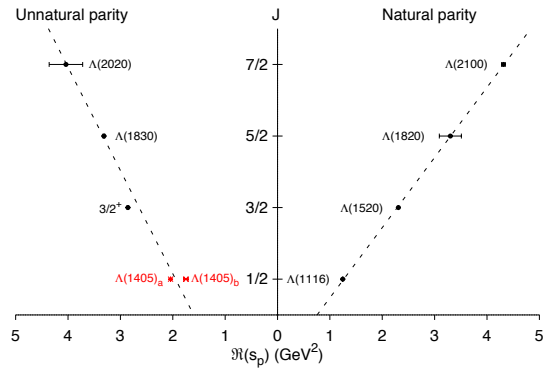}}} \label{fig:poles0rega}} 
\subfigure[\ $\Sigma$ resonances.]{
\rotatebox{0}{\scalebox{0.3}[0.3]{\includegraphics{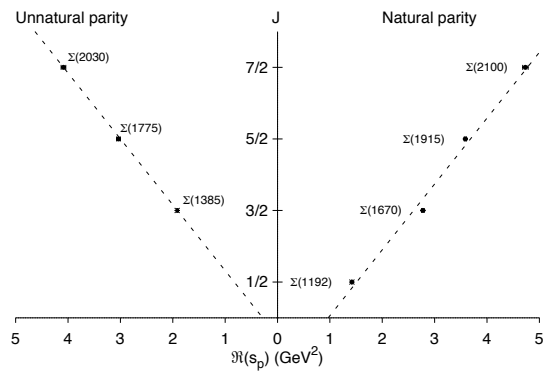}}} \label{fig:poles1rega}} 
\caption{Chew--Frautschi plot for the leading $\Lambda$  and $\Sigma$ Regge trajectories.
Dashed lines are displayed to guide the eye.} \label{fig:polesreg1}
\end{figure}

The Regge trajectory,  $\alpha (s)$, is an 
analytical function with right-hand discontinuities determined by 
unitarity.  Resonance poles, $s_p$, fulfill the conditions  $\Re\left[ \alpha (s_p)\right]=J$
and $\Im\left[ \alpha (s_p)\right]=0$. It is customary to plot $J$ \textit{vs.}~$\Re ( s_p )$
(Chew--Frautschi plot \cite{CFplot}),
\textit{i.e.}~the projection of the real part of the Regge trajectory onto the ($\Re ( s_p )$,$J$) plane.
Figure \ref{fig:polesreg1} shows the Chew--Frautschi plot for the
$\Lambda$ and $\Sigma$ leading Regge trajectories. The dashed lines are depicted to guide the eye.
We note that each line contains two nearly  degenerate Regge trajectories
corresponding to different signatures, \textit{e.g.}~the $I^\eta=0^+$ 
trajectory  in Fig.~\ref{fig:poles0rega} 
contains the $\Lambda (1116)$ and the $\Lambda (1820)$ while 
$\Lambda (1520)$ and $\Lambda (2100)$ lie on another trajectory with signature $\tau=-1$. 
In principle, trajectories with odd and even signatures are different. However,
the difference is due to exchange forces which in this  case appear to 
be weak  making the trajectories nearly degenerate \cite{PDBCollins,Gribov,Chiu}.
In the following we will treat these states 
as if they were part of the same Regge trajectory.
In Fig.~\ref{fig:polesreg1}, the linear alignment of 
$\Lambda$ and $\Sigma$ resonances is apparent.
This is common to ordinary (three-quark) baryons \cite{Faustov15,quarklinear,mesonregge}.
Inspecting the real part of the leading $0^-$ trajectory shown in Fig.~\ref{fig:polesreg1} 
we observe that both $\Lambda (1405)_a$ and $\Lambda (1405)_b$  
states could be attributed to the trajectory, but only one can belong to it. 
In principle, the pole that does not belong to the $0^-$ leading trajectory 
could be either an ordinary three-quark state or a nonordinary state.
If it were a three-quark state it should lie on a daughter Regge trajectory that has to be, approximately, 
parallel to the leading trajectory.
However, this second pole cannot belong to a daughter Regge trajectory 
because, if that were the case, the daughter Regge trajectory would overlap the leading trajectory.
Hence, at least one of the $\Lambda (1405)$ states is a nonordinary state, 
\textit{i.e.}~its composition should be  different from an ordinary three-quark baryon. 

\begin{figure}
\centering
\subfigure[\ $\Lambda$ resonances.]{
\rotatebox{0}{\scalebox{0.3}[0.3]{\includegraphics{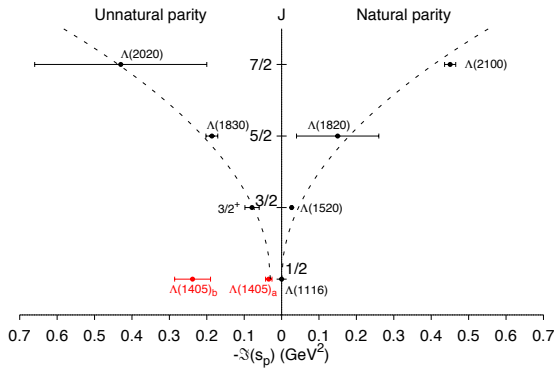}}}  \label{fig:poles0regb}}
\subfigure[\ $\Sigma$ resonances.]{
\rotatebox{0}{\scalebox{0.3}[0.3]{\includegraphics{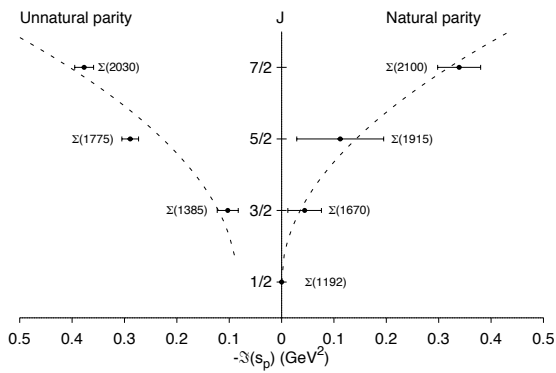}}}  \label{fig:poles1regb}}
\caption{Projections of the leading $\Lambda$  and $\Sigma$ Regge trajectories
onto the ($-\Im (s_p)$, $J$) plane. 
Dashed lines are displayed to guide the eye.} \label{fig:polesreg2}
\end{figure}

It is, in principle, possible that neither of the $\Lambda (1405)$ poles belong to the $0^-$ leading trajectory.
To further address this question,  in Fig.~\ref{fig:polesreg2} we  plot  $J$ \textit{vs.}~$-\Im ( s_p )$.  
It is apparent that both the $\Lambda$ and the $\Sigma$ trajectories follow a square-root-like behavior 
implied by unitarity that implies a relation between the phase-space volume and resonance widths \cite{Gribov}. 
The $\Lambda$ and the $\Sigma$ leading trajectories correspond to ordinary baryons
as indicated by the linear behavior in the Chew--Frautschi plot (Fig.~\ref{fig:polesreg1}). 
We find that all of these trajectories also follow a square-root-like behavior 
when the $J$ \textit{vs.}~$-\Im ( s_p )$ plot is considered.
Hence, we conclude that the Regge trajectory of ordinary baryons should follow square-root-like 
behavior in the $J$ \textit{vs.}~$-\Im ( s_p )$ plot.
Inspecting Fig.~\ref{fig:poles0regb} one concludes that $\Lambda (1405)_a$ appears on the $0^-$ 
Regge trajectory of ordinary,  three-quark, states while the  $\Lambda (1405)_b$ 
is a candidate for a new nonordinary baryon  resonance.  
In the following we summarize results of a quantitative analysis. 

To assess the model dependence of these conclusions we choose three alternative 
parametrizations of the Regge trajectory.
We define \cite{Chiu,Mandelstam}:
\begin{equation}
\alpha(s) = \alpha_0 + \alpha' s + i\: \gamma \: \rho(s,s_t) \: , \label{eq:regge}
\end{equation}
where $\alpha_0$, $\alpha'$, $\gamma$ and  $s_t$ can be obtained by fitting the poles $s=s_p$ to
$\Re \left[ \alpha(s_p) \right]= J$ and $\Im \left[ \alpha(s_p) \right]= 0$. 
For $ \rho(s,s_t)$, we use, 
\begin{alignat}{2}
 i\: \rho_{A}(s,s_t) = & i\: \sqrt{s-s_t} \: ,\\
 i\: \rho_{B}(s,s_t) = & i\: \sqrt{1-s_t/s}  \: , \\
 i\: \rho_{C}(s,s_t) =& \frac{s-s_t}{\pi}\int_{s_t}^\infty \frac{\sqrt{1-s_t/s'}}{s'-s_t}  \frac{ds'}{s'-s}  \notag \\
=&\frac{2}{\pi}\frac{s-s_t}{\sqrt{s(s_t-s)}}\arctan \sqrt{\frac{s}{s_t-s}} \: .
\end{alignat}
Model $C$ is the analytic continuation of the phase space (dispersive approach) 
where $\alpha_0$ and $\alpha'$ are the subtraction constants.
It is motivated by the relation between the imaginary part of the Regge 
trajectory and the width of the resonances \cite{Gribov}.
Models $A$ and $B$ are  alternative phenomenological parametrizations. 
Model $B$ should not be trusted on the left-hand cut that should not be present in $\alpha (s)$. 
For each model we fit the $0^+$, $1^+$, and $1^-$ trajectories that we use as benchmarks.
For the $0^-$ trajectory we fitted the three trajectories  
depending on which of the two $\Lambda(1405)$ poles is included to lie on the trajectory. 
We  refer to this trajectory as $0^-_{a(b)}$ when  $\Lambda (1405)_{a(b)}$ is included or as 
$0^-_c$ when neither pole is included.
To obtain the parameters and their uncertainties we proceed as follows. 
First, we randomly choose values for the pole positions  $s_p$ by sampling 
a Gaussian distribution according to the uncertainties given in Table \ref{tab:poles}. 
We  use the least-squares method to fit the trajectory parameters, Eq.~(\ref{eq:regge}), 
by  minimizing the distance $d$ between the trajectory $\alpha(s)$ evaluated at the complex 
pole position $s=s_p$ and the real angular momenta $J$, 
\begin{equation} 
d^2 = \sum_{poles} \{ \:  \left[ J- \Re \alpha(s_p) \right]^2 +  \left[ 0- \Im  \alpha(s_p) \right]^2 \: \}\: .
\end{equation} 
The procedure is repeated, each time  obtaining a new set of trajectory parameters. 
The expected value of each parameter is computed as the mean of the 
$10^4$ samples and the uncertainty is given by the standard deviation. 
The results are summarized in Table \ref{tab:regge}.

\begin{table}
\caption{Fitted parameters of the leading Regge trajectories as defined in Eq.~(\ref{eq:regge}).  
The parameter $\gamma$ has units of GeV$^{-1}$ for model $A$ and
is dimensionless for models $B$ and $C$.} \label{tab:regge}
\begin{ruledtabular}
\begin{tabular}{cccccc}
Model&$I^\eta$ & $-\alpha_0$ &$\alpha'$ (GeV$^{-1}$) & $\gamma$& $s_t$  (GeV$^2$)\\
\hline
$A$&$0^-_a$ & $3.3(1.5)$& $1.68(43)$ &$0.56(50)$ & $2.44(65)$\\
&$0^-_b$ & $2.19(76)$& $1.37(24)$ &$0.35(31)$ & $1.2(1.1)$\\
&$0^-_c$ & $3.4(1.9)$& $ 1.70(58) $ &$ 0.62(48)$ & $ 2.60(82) $\\
&$0^+$ & $1.25(58)$&$1.09(12)$ & $0.37(19)$&$2.63(78)$ \\
&$1^-$& $0.317(86)$&$0.924(27)$ &$ 0.236(21)$ &$1.79(14)$ \\
&$1^+$& $0.858(64)$&$0.913(19)$ & $0.113(27)$&$1.47(45)$ \\
\hline
$B$&$0^-_a$ &$3.5(1.7)$&$ 1.75(52) $ & $ 1.02(77)$&$ 2.43(58) $ \\ 
&$0^-_b$ &$2.6(1.3) $&$ 1.50(38) $ & $ 0.81(67) $&$ 1.5(1.1) $ \\
&$0^-_c$ & $3.4(1.9)$& $ 1.73(59) $ &$ 1.17(76)$ & $ 2.64(69)$\\
&$0^+$ & $1.22(86)$&$ 1.09(20)$ & $0.52(35) $&$ 2.08(94)$ \\
&$1^-$& $0.41(13)  $&$0.953(39)$ & $0.482(48)$&$ 1.92(13)$ \\
&$1^+$&  $0.855(88)$&$ 0.913(23)$ & $ 0.203(57) $&$ 1.6(1.1) $ \\
\hline
$C$&$0^-_a$ & $3.9(2.1) $&$ 1.69(41) $ & $ -2.2(2.7) $&$ 2.92(87) $ \\
&$0^-_b$ &$2.21(86)$&$ 1.30(22) $ & $ -0.7(1.1) $&$ 1.4(1.2)$ \\
&$0^-_c$ & $3.1(2.1)$& $ 1.57(58) $ &$ -1.4(1.4) $ & $ 2.78(80) $\\
&$0^+$ & $1.54(85)$&$ 1.10(12) $ & $ -1.3(1.1)$&$ 3.06(91) $ \\
&$1^-$& $0.26(21) $&$0.861(32) $ & $-0.471(63)$&$1.91(26)$ \\
&$1^+$&$1.09(22) $&$ 0.944(32) $ & $ -0.47(29) $&$ 2.87(50) $ \\
\end{tabular}
\end{ruledtabular}
\end{table}

The canonical  values of the intercept  $\alpha_0$ and slope $\alpha'$ can be found in
\textit{e.g.}~Refs.~\cite{PDBCollins,HighEnergy}.
Typically, these parameters are obtained from fits to the real part of the trajectory only 
\textit{i.e.}~using the relation  $J = \bar{\alpha}_0 + \bar{\alpha}' M^2$ with 
$M$ being the Breit--Wigner mass of the resonance.  
The canonical values are
$\bar{\alpha}_0 \simeq -0.6$ and $\bar{\alpha}' \simeq 0.9$ GeV$^{-2}$ for the $0^+$ trajectory and
$\bar{\alpha}_0 \simeq -0.8$ and $\bar{\alpha}' \simeq 0.9$ GeV$^{-2}$ for the $1^+$ trajectory \cite{PDBCollins}.
These yield good results, for example,  when applied to backward  $K^+p\to K^+p$ reaction at high energy, 
where hyperon exchange far from threshold dominates  the cross section \cite{HighEnergy}.
The  intercepts  $\alpha_0$ for the $0^+$ and $1^+$  trajectories were also obtained
in Ref.~\cite{Berger1969} by fitting the high-energy kaon backward scattering data 
(with $\alpha'$ fixed to $1$ GeV$^2$)  yielding $\alpha_0=-1.24$ or $-1.15$  for the $0^+$ 
trajectory and $\alpha_0=-0.9$ or $-0.8$ for the $1^+$. 
If we  limit our analysis to the real parts parametrized by linear functions  we obtain 
$\bar{\alpha}_0\simeq -0.74$ and $\bar{\alpha}' \simeq 0.98$ GeV$^{-2}$ for  $0^+$ and
$\bar{\alpha}_0\simeq -0.89$ and $\bar{\alpha}' \simeq 0.92$ GeV$^{-2}$ for  $1^+$.
The results of our analysis (Table \ref{tab:regge}) 
obtained by  fitting trajectory parameters in the resonant region using latest values of the pole positions 
(Table \ref{tab:poles})  are consistent with the earlier fits. 

\begin{figure}
\begin{center}
\rotatebox{0}{\scalebox{0.3}[0.3]{\includegraphics{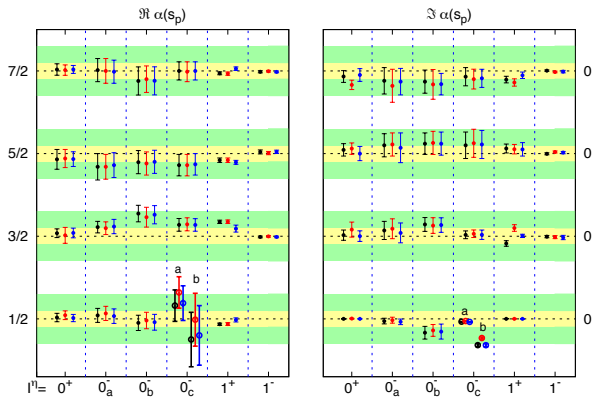}}}
\caption{Consistency check. Left plot shows $\Re \left[ \alpha(s_p) \right]$
for each one of the six fitted Regge trajectories 
--\textit{i.e.}~$0^+$, $0^-_a$, $0^-_b$,$0^-_c$, $1^+$, and $1^-$ (see text)--,
computed at the poles of the resonances ($s_p$) for models $A$ (black), $B$ (red), and $C$ (blue). 
The result should be equal to the corresponding angular momentum $J$ (vertical axis) for a given resonance. 
Right plot shows the same calculation for $\Im \left[ \alpha(s_p) \right]$, which should be equal to zero. 
For the $0^-_c$ columns, the lowest points represent the 
$\alpha(s)$ predictions of the $0^-_c$ fit at $\Lambda (1405)_a$ and $\Lambda (1405)_b$ poles.
The yellow (green) bands represent up to 0.1 (from 0.1 to 0.3) 
deviation from the label in the vertical axis.} \label{fig:reggetest}
\end{center}
\end{figure}

To further assess the quality of the fits shown in Table  \ref{tab:regge} 
we perform the following consistency check. 
For a given model at a given $s$, $\alpha(s)$ is 
computed as the mean value of the $10^4$ fits performed to obtain the parameters listed in Table \ref{tab:regge}.
At the location of the poles, $s=s_P$ one  should  find, 
within fit uncertainties, $\Im \left[ \alpha(s_p)\right]=0$   
and $\Re  \left[ \alpha(s_p)\right]=J$. 
The extent to which these conditions are satisfied is depicted in Fig.~\ref{fig:reggetest}. 
For the  $0^+$ and $1^-$ trajectories, for all models the agreement is excellent. 
The  $1^+$ trajectory shows the superiority of the dispersive model $C$. 
It recovers $\Im \left[ \alpha(s_p)\right]=0$ for all of the poles  while models $A$ and $B$ do not. 
Model $C$ has some difficulty to recover $\Re  \left[ \alpha(s_p)\right]=J$ for $J=3/2$ and $5/2$ resonances, 
but it still provides a better description than models $A$ and $B$. 
The disagreement between the Regge model and the data 
is most likely due to the small uncertainty in the
pole parameters, which, as discussed in \cite{FR2015}, 
may have been underestimated for some resonances due to systematics in the data. 
For  the $0^-_a$ trajectory all of the models reproduce $\Re  \left[ \alpha(s_p)\right]=J$ 
and $\Im \left[ \alpha(s_p)\right]=0$ although there is certain tension at $J=3/2$ 
for the real part of the Regge trajectory.  This is expected after inspection of  Fig.~\ref{fig:poles0rega}.
On the other hand, for all the models,  the fitted $0^-_b$ trajectory fails to fulfill the conditions 
$\Im \left[ \alpha(s_p)\right]=0$ and the condition $\Re  \left[ \alpha(s_p)\right]=J$ 
is violated for the $3/2^+$ state.   
It also fails to reproduce the $\Im \left[ \alpha(s_p)\right]=0$ condition for $\Lambda (1405)_b$. 
Fits to $0^-_c$ have no information about the $\Lambda (1405)$ states and 
we can check if we obtain $\Re  \left[ \alpha(s)\right]=1/2$ and  $\Im \left[ \alpha(s)\right]=0$ 
at either of the two $\Lambda (1405)$ poles.
We find that the $0^-_c$ fit  provides the correct result for the $\Lambda (1405)_a$ 
state but not for the $\Lambda (1405)_b$ where the condition 
$\Im\left[  \alpha (s_p)\right]=0$ is not satisfied. 
The consistency check supports the qualitative results obtained from
Figs.~\ref{fig:polesreg1} and \ref{fig:polesreg2} inspection.

We find a consistent picture for the leading hyperon Regge trajectories. 
Using the  $\Sigma$ and $0^+$  trajectories as the benchmark for the ordinary, three-quark states 
we find that one of the $\Lambda (1405)$ poles, denoted here as $\Lambda(1405)_a$, which has pole mass
$1429-i \: 12$ MeV, belongs to the $0^-$ leading Regge trajectory and 
therefore is  most  likely dominated by the ordinary three-quark configuration. The $\Lambda (1405)_b$  pole, 
located at $1325-i\: 90$ MeV, does not belong to either the $0^-$ leading Regge trajectory or a close by daughter.  
Hence, $\Lambda (1405)_b$ does not seem to fit the common pattern of a linear Regge trajectory   
of  known three-quark  hyperons possibly indicating  its nonordinary nature. 
This result is consistent with quark-diquark model expectations 
which find only one of the $\Lambda$(1405) states \cite{Santopinto15,Faustov15},
large $N_c$ calculations obtaining a three-quark state in the $\Lambda(1405)$ region \cite{LargeNc},
and  with lattice QCD calculations obtaining either a three-quark \cite{Engel13} 
or a $\bar{K}N$ \cite{Hall2015} state.  Further studies should assess if the nature of $\Lambda (1405)_b$
is that of a pentaquark or a molecular state, although the last interpretation is favored
by the literature \cite{twopoles,Mai2015,molecule}.

\begin{acknowledgments}
We thank Ra\'ul Brice\~no,  Jos\'e Goity, Michael Pennington, and Alessandro Pilloni for useful discussions.
This material is based upon work supported in part by the U.S.~Department of Energy, Office of Science, 
Office of Nuclear Physics under contract DE-AC05-06OR23177. 
This work was also supported in part by the U.S.~Department of Energy under Grant DE-FG0287ER40365, 
National Science Foundation under Grants PHY-1415459 and PHY-1205019, and IU Collaborative Research Grant.
\end{acknowledgments}


\end{document}